\documentclass[letterpaper]{article}

\usepackage[T1]{fontenc}
\usepackage[colorlinks=true,urlcolor=blue,citecolor=blue,linkcolor=blue]{hyperref}
\usepackage{amssymb,amsthm,amsmath,amsfonts}
\usepackage{graphicx,mathptmx}
\usepackage[pdftex,dvipsnames,usenames]{xcolor}
\usepackage[colorlinks=true,urlcolor=blue,citecolor=blue,linkcolor=blue]{hyperref}
\usepackage{comment,bm}
\usepackage{physics, amsmath, amssymb, amsthm, units, dsfont, bm}
\usepackage[shortlabels]{enumitem}
\usepackage{soul}

\usepackage{geometry}
\usepackage{setspace}

\usepackage[style = chem-acs]{biblatex}
\usepackage{graphicx}
\usepackage{float}
\newfloat{scheme}{htbp}{los}
\floatname{scheme}{Scheme}
\floatname{chart}{Chart}
\newfloat{graph}{htbp}{loh}

\usepackage{chemformula} 
\usepackage[version = 4]{mhchem} 

\usepackage{authblk}
\author[1]{Nidhin Prasannan}
\author[1]{Konstantinos Mourzidis}
\author[1]{Vishwas Jindal}
\affil[1]{Universit\'e de Toulouse, INSA-CNRS, LPCNO, 135 Av. Rangueil, 31077 Toulouse, France}
\author[2]{Hanting Li}
\author[2]{Di Lin}
\author[2]{Wenchen Yang}
\author[2]{Deng Hu}
\author[2]{Liu Yang}
\affil[2]{Key Laboratory of Advanced Optoelectronic Quantum Architecture and Measurement, Ministry of Education, School of Physics and Beijing Key Lab of Nanophotonics and Ultrafine Optoelectronic Systems, Beijing Institute of Technology, Beijing 100081, China}
\affil[3]{Beijing Institute of Technology, Zhuhai 519000, China}
\author[1]{Andrea Balocchi}
\author[1]{Delphine Lagarde}
\author[1]{Pierre Renucci}
\author[2,3]{Zhiwei Wang}
\author[2,*]{Gang Wang}
\author[1,4,*]{Xavier Marie}
\affil[4]{Institut Universitaire de France, 75231 Paris, France}

\title{Entangled Telecom Photon Generation using Twisted Van der Waals Crystals}
\date{*Email: marie@insa-toulouse.fr, gw@bit.edu.cn}

\begin{document}

\maketitle

\begin{abstract}
  Nanoscale quantum light sources are essential building blocks for integrated quantum photonic systems. Here, we report a wavelength-scale entangled-photon source based on van der Waals-engineered NbOBr$_2$, and benchmark its performance for telecom-wavelength quantum light generation. By exploiting the material's second-order nonlinearity, we generate quantum-correlated photon pairs via spontaneous parametric down-conversion. We then use a 90$^{\circ}$ twisted stacking to induce quantum interference in photon-pair generation, yielding polarization-entangled photons. This approach enables tunability of the quantum optical state via control of the excitation laser polarization. We experimentally obtain entanglement fidelities exceeding 95\% for Bell states, along with a high coincidence-to-accidental ratio of $\sim$335, and a brightness approximately one order of magnitude higher than recently reported telecom sources based on transition metal dichalcogenide~(TMD) 2D materials. These results establish twisted van der Waals engineering as a powerful platform for highly tunable, high-brightness quantum light sources at telecom wavelengths.
\end{abstract}



\section{Introduction}

The generation of entangled photon pairs is a central resource for quantum communication, quantum networks, and photonic quantum information processing~\cite{scarani09, pelucchi22}. Telecom wavelength sources are particularly desirable because they align with low-loss optical fiber transmission and the current photonic infrastructure. Spontaneous parametric down-conversion (SPDC) in bulk nonlinear crystals has long been the workhorse for entangled photon generation~\cite{couteau18, ali21, kwiat95, evan18}, but scaling such sources toward integrated, compact, and tunable platforms remains an ongoing challenge.

Two-dimensional (2D) materials have recently emerged as a promising alternative platform for nonlinear and quantum optics, owing to their strong optical nonlinearities, reduced dimensionality, and the possibility of integration with photonic and electronic systems~\cite{sarkar26}. Also, the rich physics emerging from the twist-angle-stacked van der Waals~(vdW) 2D materials have stimulated many new research directions, including moiré engineering, unconventional superconductivity, and  strongly correlated many-body condensates~\cite{cao18,he21,koppens20,luojun23}. 

While second-harmonic generation (SHG) has been extensively studied in a wide range of 2D materials, the demonstration of SPDC and entangled photon pair generation has only very recently been achieved. Early demonstrations include layered transition metal dichalcogenides~(TMD) such as 3R-MoS$_2$ or WS$_2$~\cite{weissflog24,feng24}, GaSe~\cite{lu25} and rhombohedral boron nitride~\cite{lyu25,liang25}, establishing the feasibility of quantum light generation in atomically thin systems. Furthermore, owing to their reduced dimensionality, ultrathin vdW materials enable broadband photon-pair emission due to relaxed phase-matching conditions~\cite{chekova21}. More recently, the family of polar vdW materials NbOX$_2$ (X = Cl, I) has attracted significant attention, with demonstrations of efficient SPDC and polarization-entangled photon generation in NbOCl$_2$ and NbOI$_2$~\cite{guo23,kallion25,qiu24,joshi26}. In these works, the nonlinear process is strongly enhanced by excitonic resonances, enabled by pumping at photon energies above the material bandgap, typically around 400~nm, in agreement with theoretical predictions of exciton-assisted SPDC enhancement~\cite{xuan24}. 

Despite these advances, two key challenges remain. First, the demonstrated sources predominantly rely on above-bandgap excitation of these 2D semiconductors, which introduces absorption losses and potential photocarrier effects. Second, the generated photon pairs have not been directly engineered for telecom wavelengths within this NbOX$_2$ material platform, restricting their immediate applicability to quantum communication technologies. Here we explore a complementary regime based on below-bandgap excitation. We report the first demonstration of SPDC and polarization-entangled photon pair generation in niobium oxide dibromide~(NbOBr$_2$), a material previously limited to classical nonlinear optics, i.e., second-harmonic generation investigations~\cite{ye23,chen24}. Using a 725~nm laser pump, well below the optical bandgap ($\sim$1.90~eV i.e 650~nm)~\cite{tang25}, we show that efficient SPDC can be achieved without excitonic enhancement. Importantly, our approach enables, for the first time in the NbOX$_2$ family, direct generation of entangled photon pairs in the telecom spectral range with a brightness approximately one order of magnitude higher than previously reported telecom-wavelength entangled-photon sources based on TMD film and higher than TMD-based vdW metasurface sources~\cite{weissflog24,fan25}. This establishes NbOX$_2$ as a new platform for low-loss, telecom-compatible quantum light sources and demonstrates that excitonic resonance is not a prerequisite for quantum nonlinear optics in this family of 2D materials.
\section{Results}
    \begin{figure*}[t]
            \centering
            {\includegraphics[width=.95\textwidth]{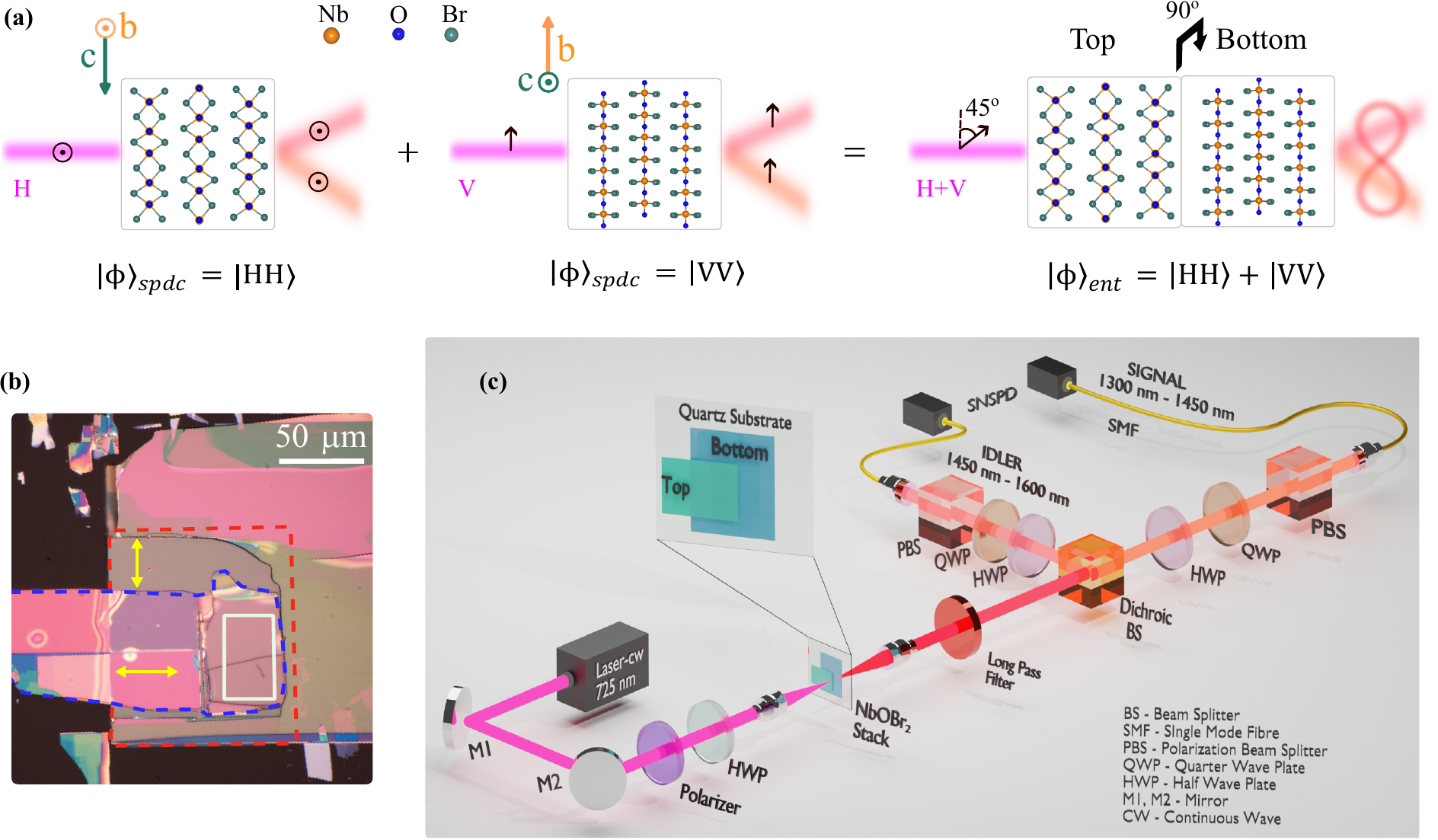}}   
            \caption{\textbf{Schematics for the generation of entangled photons from a van der Waals NbOBr$_2$ stack}. (a)~Representation of the crystal structure and axes orientation of the two NbOBr$_2$ crystals. The polar b-axis of the two crystals are orthogonal to each other due to the relative 90$^{\circ}$ twist angle between the crystals. The H and V directions are defined with respect to the laboratory frame. The orthogonal b and c arrows indicate the orientation of the crystallographic axes. (b)~Optical microscope image of the orthogonal NbOBr$_2$ stack used in the experiment. Top and bottom flakes are marked with dotted lines (blue and red, respectively). The marked yellow direction on the stacks reveals the crystal b-axis, obtained from nonlinear conversion experiments~(SHG and SPDC). The experimental region where we performed SPDC and entanglement tests is marked with a white rectangle. (c)~Schematic representation of the experimental setup utilized for the SPDC and entanglement measurements~(detailed description in Methods).
            }\label{fig:sample}
    \end{figure*}

NbOBr$_2$, part of the NbOX$_2$ family, exhibits C$_2$ symmetry, with bromine (Br) occupying the `X' position in the crystal structure, resulting in an oxyhalide compounds~(Fig.~\ref{fig:sample}a). In the context of nonlinear optics, the second-order susceptibility $\chi^{(2)}$ tensor elements of the NbOBr$_2$ crystals follow the same electric field polarization selection rules to couple the fundamental and second-order field modes~\cite{ye23}. We have fabricated vdW stacks with two NbOBr$_2$ crystals that are stacked orthogonal to each other with respect to their b-c plane~(see Fig.~\ref{fig:sample}a and Methods). Here, the b and c directions represent the polar and non-polar axes of the NbOX$_2$ crystal. In order to maximize the entangled photon pair generation rate, each layer has almost the same thickness $\sim$600~nm. Under normal incidence to the crystal b-c plane, the excitation and emission field polarizations are restricted to in-plane~(b-c) components only. The tensor elements contributing to the parametric SHG process are $\chi^{(2)}_{bbb},~ \chi^{(2)}_{bcc}$ and $\chi^{(2)}_{cbc}$ where the subscripts denote the polarization directions of the interacting fields. The first index indicates the polarization of the SHG field, while the second and the third ones the polarization direction of the incident fields. 
Among these, the $\chi^{(2)}_{bbb}$ tensor element is the largest one ~\cite{guo23, ye23}.  For the $\chi^{(2)}_{bbb}$ SPDC process, when the excitation polarization is aligned along the b-axis, the generated nonlinear fields preserve the same polarization. This is a general electric field polarization selection rule for the type-0 SPDC~\cite{boyd20}.  Accordingly, excitation of an NbOX$_2$ crystal along the b-axis results in the generation of correlated photon pairs co-polarized along the b direction,~$\big(\mathrm{b_{pump}}\rightarrow\ket{\mathrm{b_sb_i}}\big)$,~ arising from the spontaneous decay of the pump photon (s and i indices correspond to signal and idler respectively) ~\cite{weinberg70}. These polarization selection rules, linking the classical nonlinear response to the quantum regime, are experimentally verified in the following.\\

    \begin{figure*}[t]
        \centering
        {\includegraphics[width=.95\textwidth]{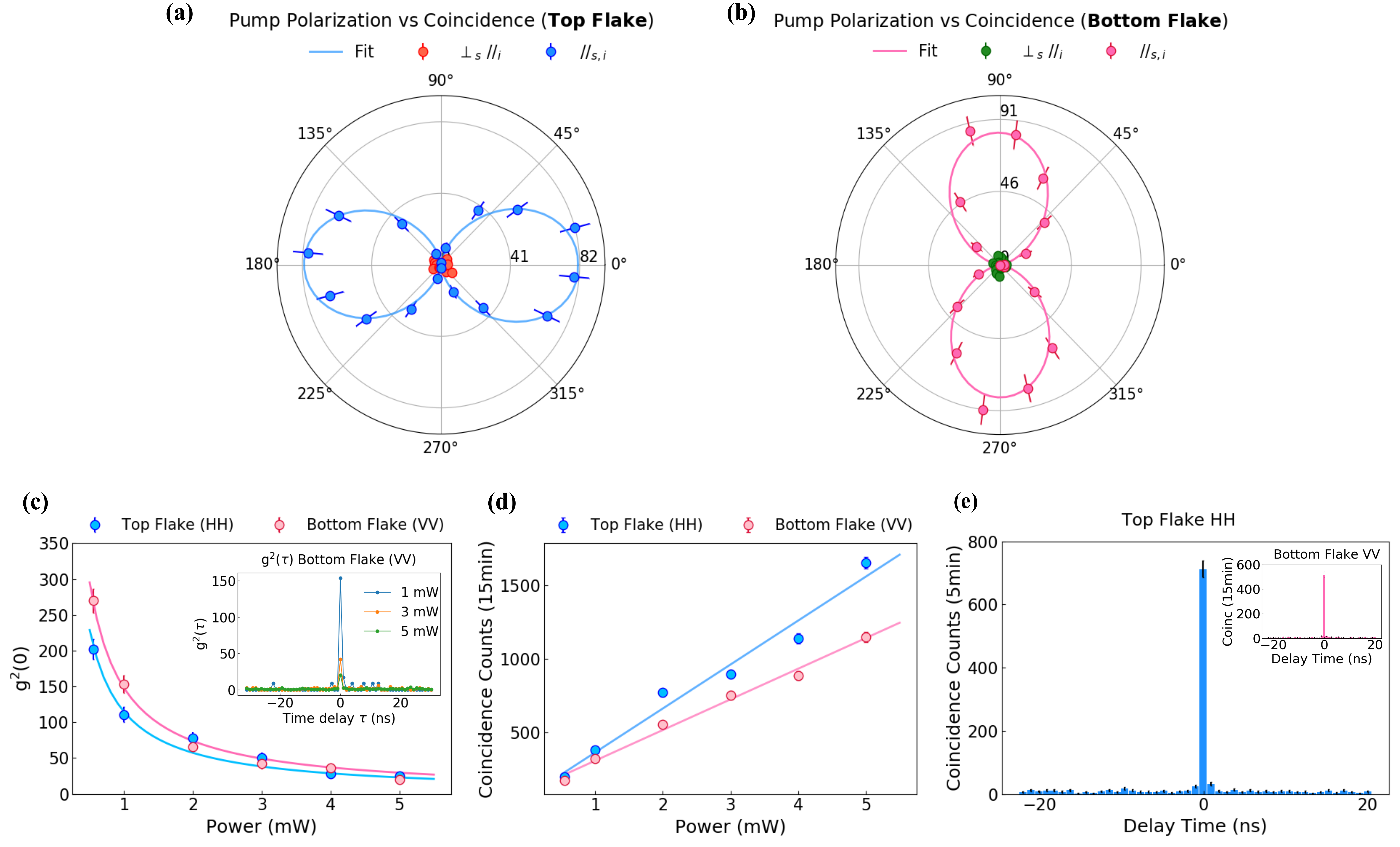}}    
            \caption{\textbf{SPDC experimental characterization from top and bottom flakes of the NbOBr$_2$ stack}. 
            (a, b) Pump polarization~(polar axis) dependent SPDC coincidence variation from the top and bottom flakes respectively, for 2 minutes acquisition. In co-polarized measurement setting~(//$_{s,i}$), the brightest SPDC signal along the horizontal and vertical pump polarization corresponds to the excitation polarization along the crystal b-axis. From both the top and bottom flake, negligible coincidences are recorded for cross-polarized setting~($\perp_s $//$_i$) irrespective of the pump polarization. (c)~Power-dependent normalized $g^{(2)}(0)$ obtained from both the top and bottom flakes. (inset) Normalized $g^{(2)}(\tau)$ from the bottom flake for three different powers. (d)~Coincidence counts from the top and bottom flake for different input pump powers (15 minutes integration time). Poissonian error bars are used in all datasets.(e)~Raw coincidence histogram data~(CAR $\sim$111) from the top flake for 2~mW of input pump power from a co-polarized~(pump and SPDC HWP's at horizontal polarization) measurement setting. Inset: histogram data~(CAR $\sim$102) from the bottom flake for vertical co-polarized settings with similar pump power.       }\label{fig:spdc}
    \end{figure*}
We first evaluate the nonlinear characteristics of individual NbOBr$_2$ 2D flakes. The crystallographic axes 
orientation and nonlinear tensor elements of the flakes can be revealed from both SHG and SPDC-based 
experimental schemes. In our measurements, the SHG process involves the excitation of a single NbOBr$_2$ 
crystal with a fundamental wavelength of 1450~nm and detection of the second harmonic at a wavelength of 
725~nm. This process is a rather direct method to determine the susceptibility~($\chi^{(2)}$) tensor 
elements of the nonlinear 2D sample due to the strong classical SHG output signal~\cite{heinz13,gang15}. We 
performed a polarization-resolved SHG measurement on individual NbOBr$_2$ flakes to identify the relevant 
crystal axes information~(see Sec.~IA of SI for experimental details). During the SHG measurements, both the 
fundamental and second harmonic polarization states are rotated in a co-polarized manner. The corresponding 
normalized SHG light intensities are plotted in the supplementary figures~(Fig S1.~b,c). For 
NbOBr$_2$, the polarization alignment with maximum SHG intensity specifies as well the crystal axis 
orientation for a strong SPDC process, which is measured to be $\chi^{(2)}_{bbb}$. The 
crystallographic polar b-axes retrieved through the SHG signal for the top and bottom flakes are marked in Fig.\ref{fig:sample}b.

Next, we use polarization-dependent SPDC process to probe the specific $\chi^{(2)}$ tensor elements. 
Fig.~\ref{fig:spdc}a,b shows the variation of the SPDC coincidence counts as a function of the pump 
polarization angle from the overlapped area of the two flakes for 2~mW excitation power. The laser wavelength 
is 725~nm, i.e. photon energy below the material optical gap energy~\cite{chen24,tamang25}. During the 
measurement, the signal and idler relative polarization axes are fixed in either co-polarized~$\big($//
$_{s,i}\rightarrow(0^{\circ}-0^{\circ}$ and $90^{\circ}-90^{\circ})\big)$ or cross-polarized~$\big(\perp_s$//
$_i\rightarrow(0^{\circ}-90^{\circ}$ and $90^{\circ}-0^{\circ})\big)$ configurations. 
    \begin{figure*}[t]
        \centering
        {\includegraphics[width=.85\textwidth]{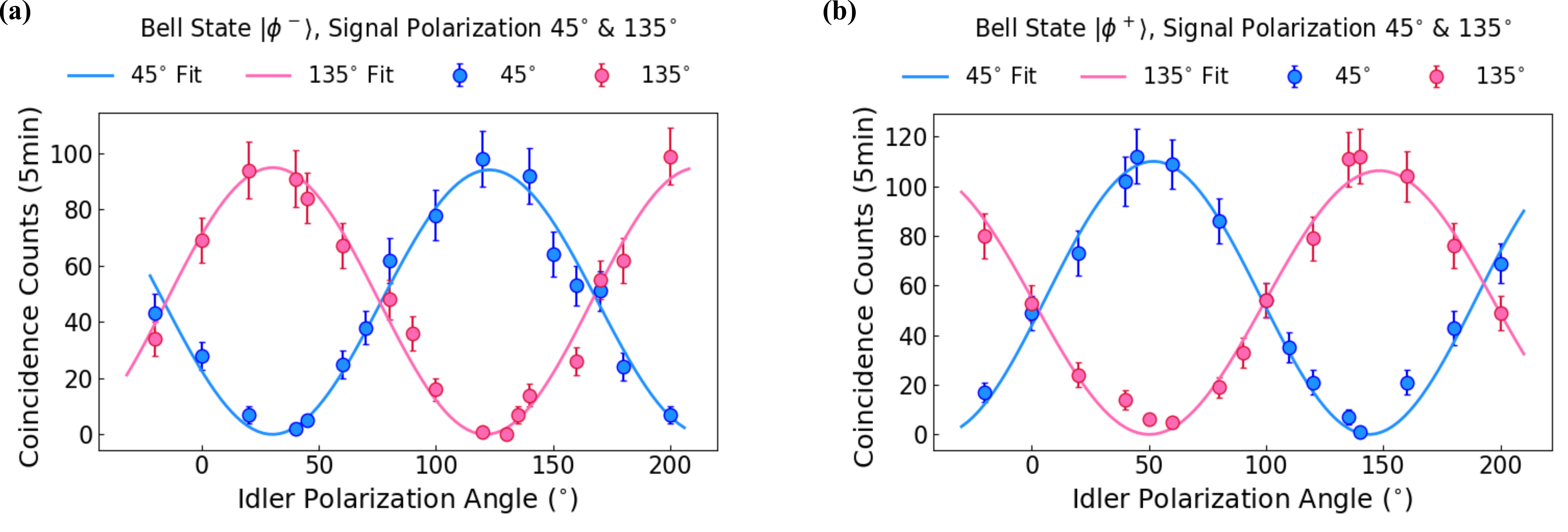}}    
            \caption{\textbf{Two photon polarization entanglement correlation}. (a)~Sinusoidally varying coincidence rate for the state $\ket{\Phi^-}$, with fixed measurement settings (D and A) at signal photon path and varying idler photon measurement basis. (b)~Similar measurements for the state $\ket{\Phi^+}$. Maximum and minimum raw coincidence rates yield the highest entanglement visibility of $\sim$98\% for the states. The integration time is 5 minutes. }\label{fig:davisibility}
    \end{figure*}
For co-polarized waveplates settings on the SPDC path, 
coincidence counts as a function of the pump polarization angle brings up a two-lobe pattern 
from top and bottom flakes and reproduces, as expected, the SHG characteristics. In 
Fig.~\ref{fig:spdc}a, we observe a maximum coincidence for the horizontal-horizontal~
(//$_{s,i}$) setting on the signal-idler side and a horizontal pump polarization, which 
corresponds to the $\chi^{(2)}_{bbb}$ tensor component of the top flake. Similarly, for 
the bottom flake co-polarized vertical~(//$_{s,i}$) setting for the SPDC and vertical pump polarization, maximize the coincidence counts. For both flakes, the b-axis aligned polarization of the analyzers and pump maximizes the coincidence, and b-axis aligned analyzers but an orthogonal pump minimizes the coincidence. The agreement on the polarization axes of the SHG and SPDC results confirms the orientation of the two flakes. Additionally, we have observed an insignificant coincidence count rate when the analyzer waveplates are in horizontal-vertical cross-polarized position~($\perp_s$//$_i$).

We have then investigated non-classical coincidence and strong photon number correlations from these vdW SPDC sources by performing second-order correlation~$g^{(2)}$ measurements.  Each NbOBr$_2$ layer generates highly correlated photon pairs $\ket{HH}$ or $\ket{VV}$ for horizontal~(H) and vertical~(V) pump polarization directions, respectively, along the crystal axis. Time-correlated photon arrivals are analyzed using a fiber-integrated Hanbury Brown-Twiss setup~\cite{ht56}. At zero delay, normalized $g^{(2)}(\tau=0)$ peaks pertain to the signature of bunched photon arrival, which also confirms the simultaneous photon pair events from the flakes~\cite{weinberg70}. This experiment is repeated individually on the top and bottom NbOBr$_2$ flake with fixed pump polarization along the crystal axis for maximum coincidence. The corresponding power-dependent normalized $g^{(2)}(\tau)$ for the bottom flake is given for different input power values~(see inset of Fig.~\ref{fig:spdc}c). We observe an inverse power dependence of normalized $g^{(2)}(0)$ from both flakes, further supporting SPDC-enabled correlated photon pair emission from the NbOBr$_2$ crystals~\cite{kitaeva21,ivanova06}.

In the low-power regime, the SPDC coincidence events counts should increase linearly with the strength of optical excitation~\cite{brida00}. Fig.~\ref{fig:spdc}d uncovers this power-dependent linear response of the coincidence counts for each flake. Experimental SPDC rates show a brightness of 0.85~Hz at 2~mW pump power, which is lower-bounded by the overall efficiency of the experimental arrangement (see SI). Counts are collected from the overlapping area, with pump polarization~(H and V) aligned with the b-axis of the individual flakes. The slight deviation in the count rates~(in Fig.~\ref{fig:spdc}d) is due to the thickness difference~($\sim$600~nm and $\sim$546~nm for the top and bottom flakes, respectively). Since we expect no correlated events except zero delay from the SPDC process histogram, the ratio of actual SPDC coincidences over random events benchmarks the quality of the source in terms of coincidence-to-accidentals~(CAR). Figure~\ref{fig:spdc}e shows the coincidence histogram data from the top flake with pump power of 2~mW, corresponding to CAR $\sim$111~(see inset for similar results~(CAR $\sim$102) for the bottom flake). At reduced pump powers, we observe even higher CAR values, exceeding 335 at an input power of 0.55~mW (see Sec.~IIA and Fig.~S3 in SI for the power dependence of the CAR). The sharp peak with strongly suppressed accidental coincidences in the histogram indicates a high signal-to-noise ratio for the NbOBr$_2$-based telecom SPDC sources. 

Note that a moderate background noise is observed in the coincidence histogram when we use only the 1100~nm long pass filter in the SPDC detection path, due to fluorescence from the NbOBr$_2$ crystals. To confirm this, we performed photoluminescence (PL) measurements on the flakes and observed a weak but broadband emission extending up to $\sim$1300~nm (see Sec.~IB in SI for details of the PL measurements). Accordingly, in all the measurements presented in this paper, we used an additional filter~(1300~nm) to suppress the PL background which results in an overall improved CAR for the NbOBr$_2$ SPDC detection. These values match with recently developed vdW metasurface SPDC sources and are higher than the TMD-only sources with similar operational wavelength and power regime~\cite{weissflog24,fan25}.\\

    \begin{figure*}[t]    
        \centering
        {\includegraphics[width=0.99\textwidth]{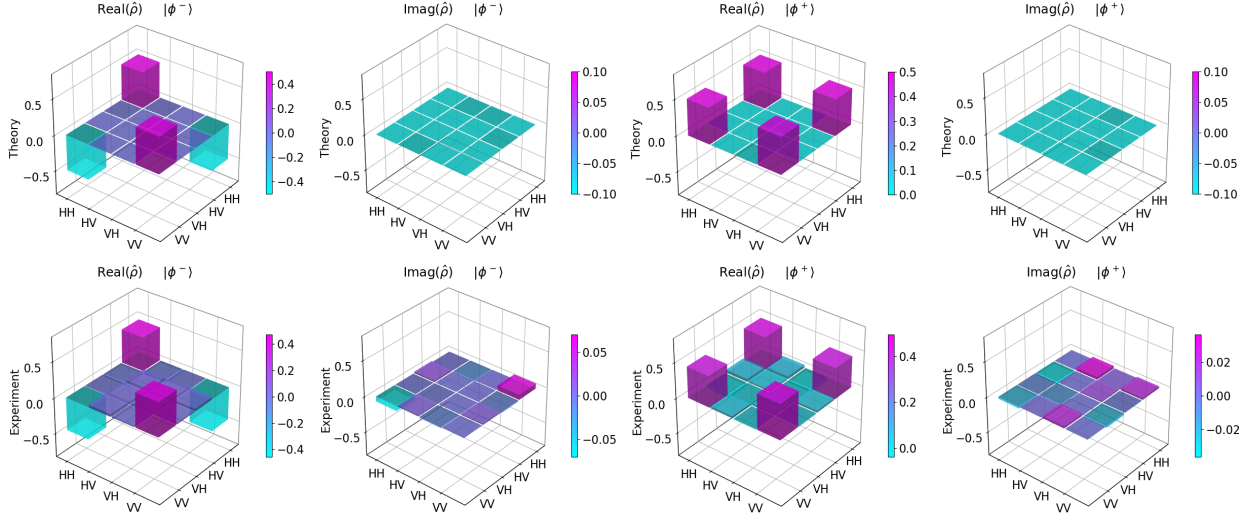}}    
            \caption{\textbf{Quantum state tomography}. (top) Theoretically expected density matrix plots for the two Bell states,~$\ket{\Phi^-}=\frac{1}{\sqrt{2}}\big(\ket{HH}-\ket{VV}\big)$ and $\ket{\Phi^+}=\frac{1}{\sqrt{2}}\big(\ket{HH}+\ket{VV}\big)$. (bottom)~ Experimentally reconstructed density matrices for states $\ket{\Phi^-}$ and $\ket{\Phi^+}$ for pump polarizations along $\pm45^{\circ}$.
        }\label{fig:tomography}
    \end{figure*}
Analogous to bulk implementations and previous approaches, we realize the
scheme using twisted stacking of NbOBr$_2$ vdW layers, generating 
polarization entangled photon pairs~(Fig.~\ref{fig:sample}
a,b)~\cite{kwiat99}. As discussed above, excitation along a single 
crystal axis leads to SPDC with a well-defined polarization state. An 
excitation optical field with components along both crystal axes (at 
45$^{\circ}$ with respect to the H direction) can simultaneously drive 
nonlinear conversion in the two overlapping layers. In our entanglement 
experiment, we exploit the same parametric down conversion process~$
\big(\chi^{(2)}_{bbb}\big)$ from the two 2D layers. When the pump is 
polarized at 45$^{\circ}$, both processes are excited with equal 
probability. Within the two-photon generation regime, the spontaneous 
decay of the pump photon from the two layers becomes indistinguishable, 
and quantum interference leads to the generation of a polarization 
entangled state:
    \begin{equation}
        \ket{\Phi^+}=\frac{1}{\sqrt{2}}\Big(\ket{HH}~+~\ket{VV}\Big).
    \end{equation}
Thus, maximally entangled Bell states can be directly generated in this stacked vdW architecture. This approach enables direct tuning of the generated quantum state within a single device, where the twist angle and pump polarization act as control parameters to access different entangled states. Notably, the high polarization selectivity observed in Fig.~\ref{fig:spdc}a,b where the process $\mathrm{b_{pump}}\rightarrow\ket{\mathrm{b_sb_i}}$ strongly dominates over $\mathrm{b_{pump}}\rightarrow\ket{\mathrm{b_sc_i}}$ and $\mathrm{b_{pump}}\rightarrow\ket{\mathrm{c_sb_i}}$ suppresses undesired channels, thereby enhancing the quantum interference visibility and the entanglement fidelity. We experimentally probe this mechanism by generating distinct Bell states through control of the pump polarization in the orthogonally stacked bilayer.

We now pump the overlapping area of the cross crystal stack~(marked in Fig.~\ref{fig:sample}b) in diagonal~(45$^{\circ}$) laser polarization with reference to the crystal b-axes, which simultaneously allows the SPDC decay process on both flakes with equal pump strength. We keep a similar pump power for the entanglement experiment (2~mW) as in the SPDC characterization. For the exact diagonal polarization setting, we observed a slight difference in the SPDC strength and coincidence count rates due to the asymmetric flake thickness from the cross-stack spots. This effect is compensated by slightly adjusting~(typically $5^{\circ}$) the diagonal polarization angle for equal coincidence rates. Such fine-tuning also retains perfect coherence and interference, leaving finer entanglement fidelity for the generated states. Polarization entanglement created by the coherent superposition of the two spontaneous decay processes $\big(H_{pump}\rightarrow\ket{H_sH_i}$ and $V_{pump}\rightarrow\ket{V_sV_i}\big)$ will be manifested by the strong correlation in the twin photon polarization state in different measurement bases. 

Figure~\ref{fig:davisibility}a,b exhibits polarization entanglement correlation between signal-idler photons when measured in diagonal~(D) and anti-diagonal~(A) bases with tomographic wave-plates. Here, one of the HWP is fixed (signal), and the other waveplate (idler) is swept to probe the nonlocal correlation in the twin-photon state, polarization measurements on the signal photon reveal its entangled idler state~\cite{bohm57}. Entanglement visibility values~$\Big(\mathcal{V}=\frac{I_{max}-I_{min}}{I_{max}+I_{min}}\Big)$ are extracted from the raw experimental data, showing very large average interference visibility of~($\mathrm{98.0}\pm\mathrm{1.03)\%}$ for $\ket{\Phi^-}$ and ($\mathrm{94.8}\pm\mathrm{2.56)\%}$ for $\ket{\Phi^+}$ in D-A bases. Consistently, high visibility of $\sim$94\% is also observed in the H-V bases~(see Sec.IIB Fig.~S4 of SI for H-V basis data). Note that it is possible to generate two different maximally entangled Bell states $\big(\ket{\Phi^+}=\frac{1}{\sqrt{2}}(\ket{HH}+\ket{VV})$ and $\ket{\Phi^-}=\frac{1}{\sqrt{2}}(\ket{HH}-\ket{VV})\big)$ with diagonal and anti-diagonal pump polarization settings and 90$^{\circ}$ twisted stacking, demonstrating the quantum state tunability with pump polarization. The corresponding correlation curves for both Bell states are given in Fig.~\ref{fig:davisibility}a,b.
    
Further quantification of the entanglement is obtained by quantum state tomography. We performed two-photon polarization tomography on the two Bell states with 16 measurement settings, and reconstructed the corresponding density matrices. Standard measurement bases and readily available reconstruction algorithms are used to obtain the density matrix and state reconstruction~\cite{dpwa01}. In Fig.~\ref{fig:tomography}, the experimentally reconstructed density matrix closely matches the ideal density
matrix of the $\ket{\Phi^{\pm}}$ states. The near-zero contribution from the imaginary part of the experimental density matrix and distinct peaks~$\big(\ket{HH},~\ket{VV}\big)$ of real part over cross polarization terms~$\big(\ket{HV},~\ket{VH}\big)$, validates phase stability and noise resilience in the entangled state creation from 2D layers and stacking. Overlapping with the ideal Bell states, experimental density matrix shows exceptional fidelity values  $(93.99\pm2.54)\%$ and $(92.25\pm2.89)\%$ for $\ket{\Phi^-}$ and $\ket{\Phi^+}$ respectively from the raw data~(see Sec.~IIB and Fig.~S5 of SI for tomographic data).

\section{Discussion}
Our NbOBr$_2$ source demonstrates a broadband~($\sim$150~nm) telecom entangled photon source, and the 
spectral detection range is limited by the imposed spectral filtering. The combination of longpass~(1300~nm) 
and dichroic~(1450~nm) filters restricts the bandwidth of the signal photons, even after photons are 
registered on the idler side outside this bandwidth~(see Methods). Remarkably, the NbOBr$_2$ SPDC source has 
a brightness of $\sim$1.8~Hz at 5~mW of pump power, which is an order of magnitude higher brightness~(0.36~Hz 
mw$^{-1}$) compared to the recently reported telecom wavelength SPDC source based on TMD film~(0.008~Hz 
mw$^{-1}$) and higher than TMD-based vdW metasurface sources~\cite{weissflog24,fan25}. Accidental corrected 
fidelity further improves the values to $F_{\phi^-}=(95.08\pm2.84)\%$ and $F_{\phi^+}=(95.22\pm2.55)\%$ from the previous estimate of the raw data. To our knowledge, the visibility and fidelity values of the NbOBr$_2$ source are among the highest compared to previous demonstrations of entanglement in NbOX$_2$ sources. Note that our NbOBr$_2$ sample has a stacking angle precision close to 90$^{\circ}$. Therefore, we attribute the high values of the quantum state's entanglement visibility and fidelity to the quality of the exfoliated crystals, classical background noise suppression and precision layer stacking. We also emphasize that the pump laser wavelength~(725~nm) we use not only allows the generation of photon pairs in the telecom wavelength range but also yields a drastic suppression of the generation of photo-carriers as a consequence of below material gap excitation~\cite{chen24}. Recently, vdW stacked quasi-phase matching has been achieved in NbOBr$_2$, promising enhanced nonlinearity and complex quantum state manipulation by going beyond two-layer stacking~\cite{tang25}. Theoretical proposals have shown that it is also possible to generate other Bell states $\big(\ket{\Psi^{\pm}}\big)$ with twisted tri-layer stacking~\cite{qiu24}. Environment-protected devices, such as hBN or Graphene encapsulation, can bear higher pump power towards brighter 2D quantum light sources~\cite{joshi26}. The brightness of the source can be further increased by incorporating metasurface structures with NbOBr$_2$ 2D layers~\cite{ling25}.

\section{Conclusion}
In summary, we have demonstrated vdW stack engineering and quantum light manipulation with NbOBr$_2$ 2D material. In particular, we evidenced telecom-wavelength SPDC and entangled photons under below-bandgap excitation conditions, in contrast to previous approaches that relied on above-bandgap pumping. This opens the possibility to explore bandgap independent excitation strategies~(above-bandgap, below-bandgap, and resonant) not only in the NbOX$_2$ family but more broadly in van der Waals engineered 2D materials, enabling quantum light sources across a wide spectral range. The generated quantum states exhibit high-quality polarization entanglement, with fidelities comparable to or exceeding those reported in other 2D platforms and conventional bulk sources. Moreover, the source demonstrates a brightness approximately one order of magnitude higher than previously reported telecom-wavelength 2D sources. Furthermore, such compact 2D devices eliminate the need for bulk interferometric configurations typically required for SPDC-based polarization entanglement~\cite{shalm15,evan18,juan17}. The ultra thin nature of 2D stacked sources also avoids the need of dispersion compensation elements required for bulk crystal stacked sources to achieve high quality entanglement~\cite{kwiat95,kwiat99} Finally, we outline potential strategies for further enhancing the source brightness, providing a pathway for future experimental developments.

\section{Methods}
High-quality NbOBr$_2$ single crystals were grown by a chemical vapor transport method with NbBr$_5$ as the transport agent. Nb powder, Nb$_2$O$_5$ powder, and NbBr$_5$ were used as starting materials with a stoichiometric ratio of Nb: O: Br~=~1: 1: 2. The sealed ampules were placed into a horizontal two-zone tube furnace with the starting material in the hot side (source side). Both zones were heated to 873 K in 10 h and maintained for 120 h, followed by cooling down to 583 K (source side) and 513 K (sinkside) over a period of 240 h. Finally, the furnace was cooled to room temperature with the power switched off.

The investigated sample consists of mechanically exfoliated NbOBr$_2$ flakes that are stacked orthogonally on a transparent quartz substrate using a deterministic dry transfer method~\cite{tang25,a2dm14}. Fig.~\ref{fig:sample}b displays an image of the NbOBr$_2$ orthogonal stack, obtained via a bright-field optical microscope. The thickness of each flake has been measured by atomic force microscopy~(AFM); 600~nm and 546~nm for the top and bottom flakes, respectively.

The vdW stack is pumped by a continuous-wave (CW) 725~nm laser diode, yielding degenerate photon pairs at wavelengths centered around 1450~nm. Along the pump laser path, a polarizing beam-splitter (PBS) and half-wave plate (HWP) combination sets the pump polarization with respect to the crystal axes to access more favorable, intense nonlinear tensor elements, thereby maximizing the nonlinear conversion rates. The pump laser is focused on the NbOBr$_2$ stacks with an aspheric lens~($\mathrm{N.A}=0.5$), and another aspheric lens is used to collect the transmitted SPDC light through the transparent substrate. The collection lens has an anti-reflection coating to reduce losses on the SPDC light. Nanoscale vdW sources generally have a broad emission angle, but we collect photons in collinear geometry~(see Fig.\ref{fig:sample}c). To suppress pump and photoluminescence leakage to the single-photon detectors, we used a 1100~nm and 1300~nm long pass~(LP) filter on the transmitted optical path. SPDC photons above the cutoff wavelength of the LP filter are separated into two spatial modes (signal and idler) by a dichroic beam-splitter~(BS). The dichroic will transmit photons below 1450~nm wavelength and reflect photons above 1450~nm. This essentially provides a 150~nm detection window for the signal photons. Wavelength-separated signal-idler photons are then coupled to single-mode fibres~(SMF) with an achromatic lens~($f=25$~mm). The single-mode nature of the collection strategy can provide high spatial purity and increased signal-idler correlation. Finally, SPDC photons are detected by broadband superconducting nanowire detectors with $<25\%$ detection efficiency. Single and coincidence events are recorded with a precision time-tagging unit~(PicoQuant) for further post-processing. Optimum focusing conditions and brightness are achieved by adjusting the lens positions and maximizing the counts in the time-tag unit. 

For biphoton correlation measurements~(g$^{(2)}$), SPDC light from NbOBr$_2$ flakes is coupled into a single-mode fibre and fed into a fibre beam splitter. Coincidence counts from the beam splitter output are recorded with the two detectors and time-stamped for correlation evaluation. 

For the two-photon polarization entangled state, quantum state tomography requires 16 sets of coincidence measurements in different measurement bases. A set of polarization tomographic measurement units is used on the signal-idler photon paths for two-photon polarization state tomography. This has a combination of a HWP and a quarter-wave plate~(QWP), followed by a PBS for projective polarization measurements.

\section*{Acknowledgements} 
 This work has been done with the financial support from the National Key R$\&$D Program of China~(Grant No. 2024YFA1408700), the Beijing-Tianjin-Hebei Collaborative Innovation Project (Grant No. 24YFXTHZ00140), the National Natural Science Foundation of China (Grants Nos. 12321004 and 12074033), the Beijing Natural Science Foundation (Grant No. Z210006). G.W acknowledges the use of facilities in Analysis and Testing Center at BIT. We also acknowledge research funding support from the Agence Nationale de la Recherche under the program ESR$\backslash$EquipEx+~(Grant No. ANR-21-ESRE-0025), the ANR project SOT-SpinLED, the EUR NanoX Grant No. ANR-17-EURE-0009 in the framework of the ``Programme des Investissements d'Avenir", and the Challenge R\&T CNES. We also thank Evan Meyer-Scott and Thierry Amand for insightful comments and suggestions.

\section*{Supporting information}

Additional experimental details and data figures are provided in the supplementary information. Polarization-resolved second-harmonic generation and photoluminescence methods and results from the NbOBr$_2$ crystals are described in Sec.IA, Sec.IB. Figure of merit of the NbOBr$_2$ SPDC source in terms of efficiency and CAR is provided in Sec.IIA. Polarization entanglement correlation measurements in H-V bases and tomographic measurement datasets are provided in Sec. IIB.

\printbibliography
\end{document}